\documentclass[journal,draftcls,onecolumn]{IEEEtran}
\usepackage{cite}
\ifCLASSINFOpdf
  \usepackage[pdftex]{graphicx}
\else
  \usepackage[dvips]{graphicx}
\fi
\usepackage[cmex10]{amsmath}
\usepackage{array}
\usepackage{theorem}
\usepackage[caption=false,font=footnotesize]{subfig}
\usepackage{fixltx2e}
\usepackage{url}
\usepackage{fancyhdr}

\makeatletter

{\theorembodyfont{\rmfamily}
   }
{\theorembodyfont{\rmfamily}
   }
{\theorembodyfont{\rmfamily}
   }
{\theorembodyfont{\rmfamily}
   }

\makeatother
\begin{document}

\pagestyle{fancy}
\fancyhf{} 
\fancyhead[L]{\small\textit{Accepted for publication in IEEE Wireless Communications Letters - Copyright transferred to IEEE}}

\title{A Physical Layer Secured Key Distribution \\ Technique for IEEE 802.11g Wireless Networks
\thanks{This work was supported in part by the MIUR project ``ESCAPADE''
(Grant RBFR105NLC) under the ``FIRB -– Futuro in Ricerca 2010'' funding program.}
}

\author{\IEEEauthorblockN{Marco Baldi, Marco Bianchi, Nicola Maturo, Franco Chiaraluce,\\}
\IEEEauthorblockA{DII, Universit\`a Politecnica delle Marche, 
Ancona, Italy\\
Email: \{m.baldi, m.bianchi, n.maturo, f.chiaraluce\}@univpm.it}
}

\maketitle

\thispagestyle{fancy}

\begin{abstract}
Key distribution and renewing in wireless local area networks is a crucial issue to guarantee
that unauthorized users are prevented from accessing the network.
In this paper, we propose a technique for allowing an automatic bootstrap and periodic renewing
of the network key by exploiting physical layer security principles, that is, the inherent differences
among transmission channels.
The proposed technique is based on scrambling of groups of consecutive packets and does not
need the use of an initial authentication nor automatic repeat request protocols.
We present a modification of the scrambling circuits included in the IEEE 802.11g standard which
allows for a suitable error propagation at the unauthorized receiver, thus achieving physical layer security.
\end{abstract}
\begin{IEEEkeywords} IEEE 802.11g, key distribution, physical layer security, scrambling, wireless networks. \end{IEEEkeywords}

\section{Introduction}
\label{sec:one}

An increasing interest is being devoted to physical layer security, which exploits the inherent randomness of the wireless channel to obtain information security
or, at least, to reduce the complexity of cryptographic techniques at higher layers.

The technique proposed in \cite{Abdallah2011} exploits physical layer security and automatic repeat request protocols to periodically
renew the secret key in IEEE 802.11 wireless local area networks (WLANs).
Such a solution requires that the authorized user has already been authenticated by the access point (AP).
So, it does not solve the problem of automatically bootstrapping the network, that is, automatically generating the first key and distributing it
only to the authorized users.

The IEEE 802.11 standard includes the Wi-Fi Protected Setup (WPS) protocol, which has been introduced in 2007 for simplifying the network bootstrap
and does not exploit physical layer security.
WPS, however, has some well-known weaknesses \cite{b0021}, which make its use insecure.

In this paper, we propose an alternative solution, based on physical layer security, which aims at creating a secure area around the AP, within which
the network key can be safely broadcast to the authorized users.
This allows them to automatically acquire the network key by simply entering the secure area.
Unauthorized users must be kept out of such area, whose size can be fixed by a suitable choice of the design parameters.
In the proposed protocol, a new key is obtained starting from the output of a linear feedback shift register (LFSR),
whose internal state is known only to authorized users.
The LFSR output can be used to compute the key for higher layer cryptographic protocols, like the Wired Equivalent Privacy and Wi-Fi
Protected Access protocols. The proposed algorithm works as follows:
\begin{itemize}
\item The AP encodes the current state of the LFSR into a set of $m$ consecutive packets and broadcasts them, without allowing requests for retransmissions.
All users know the encoding technique, which is described next.
\item If a user correctly receives the whole set of $m$ packets, he is able to decode the LFSR seed, from which he obtains the current key and gets access to the network.
\item If a user fails to receive one or more of the $m$ packets, he is unable to get any information on the current key, and has no access to the network.
\end{itemize}
After having acquired the LFSR seed, the authorized receiver (Bob) can exit the secure area, but he will continue to access
the network while staying within the coverage region of the AP, having its data protected by the higher level cryptographic protocol.

We suppose that Bob and the eavesdropper (Eve) use similar devices. The former, by coming close to the AP, experiences a smaller path loss and
a larger signal-to-noise ratio than the latter.
So, Bob has a high probability of receiving the whole set of $m$ packets without errors, while it is very likely that Eve will receive at
least one of the $m$ packets in error.
Actually, Eve could gain some advantage by using more powerful receivers, e.g., exploiting diversity techniques.
This would affect the design of the system parameters and, hence, the size of the secure area, but the approach remains valid.

The main issue of the proposed protocol is to find a suitable encoding of the secret seed into $m$ packets such that even a single error
prevents Eve from getting any information on it.
This condition is also known as \textit{avalanche effect} in cryptography, and we have recently shown that it
can be achieved by suitable scrambling operations \cite{Baldi2010, Baldi2011, Baldi2012}.
The most favorable condition is represented by the ``perfect scrambler'', for which a single residual bit error is sufficient to ensure that half of the information bits are in error after descrambling, with randomly distributed error positions.
We have shown that the perfect scrambler can be easily approached by a well-designed matrix scrambler.
The latter is appropriate in the case of packets with fixed size,
while, according to the IEEE 802.11 standard \cite{802.11g}, the packet size is variable.
The standard already includes a scrambler, aimed at avoiding long runs of symbols 0 or 1, that may be responsible for an incorrect synchronization.
So, we first assess if this component is also suitable for physical layer security.
We show that the IEEE 802.11g scrambler has no error propagation ability; so, we propose a simple modification of it, which 
is able to reach the same error propagation properties of the perfect scrambler.

The paper is organized as follows. In Section \ref{sec:two}, we study the scrambler and descrambler used in 802.11g under the physical layer security viewpoint.
In Section \ref{sec:three}, we describe the new proposal and its performance in terms of error propagation ability.
In Section \ref{sec:four}, we provide an example of performance for the whole scheme.
Finally, Section \ref{sec:five} concludes the paper.

\section{Scramblers in 802.11g}
\label{sec:two}

\begin{figure}[tb]
\begin{centering}
\includegraphics[width=70mm,keepaspectratio]{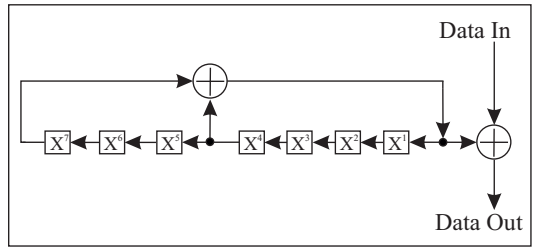}
\caption{IEEE 802.11g-a scrambler/descrambler circuit. \label{fig:scrambler_a}}
\par\end{centering}
\end{figure}

The IEEE 802.11g standard includes two types of scramblers. The first one is shown in Fig. \ref{fig:scrambler_a}, and is used when the 802.11g device is operating in 802.11a mode.
From the figure, we see that this scrambler consists of an LFSR, described by the polynomial $G(X) = X^{7} + X^{4} + 1$.
Since this polynomial is primitive, the LFSR has maximum period, equal to $127$. The output of the LFSR is added (EX-OR) to the input data, producing the output scrambled data. At the receiver side, the same LFSR is used for descrambling, thus reobtaining the original information sequence. The initial seed is randomly chosen by the transmitter but, obviously, it must be communicated to the receiver. For such purpose, the first $7$ bits of the Service field in the ERP-OFDM packet (details on the frame structure can be found in the standard \cite{802.11g}) are zero; hence, the seed is transmitted in clear and the receiver can recover it (in the absence of errors due to the channel).

\begin{figure}[tb]
\begin{centering}
\includegraphics[width=70mm,keepaspectratio]{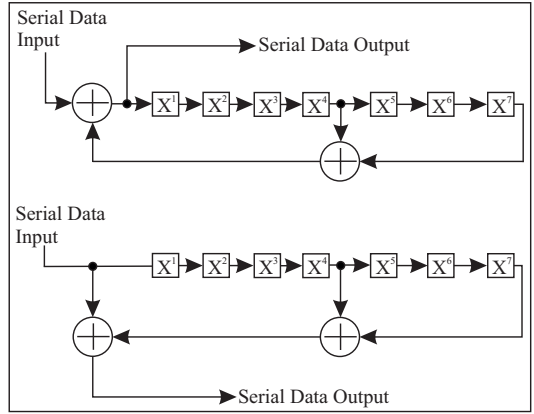}
\caption{IEEE 802.11g-b scrambler (top) and descrambler (bottom) circuits. \label{fig:scrambler_b}}
\par\end{centering}
\end{figure}

The second type of scrambler is shown in Fig. \ref{fig:scrambler_b}, and is used when the 802.11g device is operating in 802.11b mode.
In this case, the scrambler and descrambler circuits are distinct, because of a more involved structure with respect to the configuration shown in Fig. \ref{fig:scrambler_a}, where the input bits do not influence the contents of the LFSR. Moreover, the initial seed is no longer chosen randomly by the transmitter, but fixed.

As discussed in Section \ref{sec:one}, we are interested in evaluating the error propagation properties of these scramblers. Actually, this is very simple, as the scramblers' structure makes their behavior highly predictable. In particular, it is immediate to verify that a single error in the sequences at the input of the descrambler in Fig. \ref{fig:scrambler_a} produces a single error at the output. The situation is very different, and indeed close to the desired behavior from the physical layer security viewpoint, only if the error affects one of the bits in the seed. This circumstance, however, has a small probability to occur, and cannot be considered significant for the intended security purposes.

Similarly, it is also immediate to check that any combination of errors at the input of the descrambler in Fig. \ref{fig:scrambler_b} produces an error pattern at the output having three times the same weight. So, also in this case, the error propagation effect is very limited. Hence, neither the descrambler used in 802.11g-a nor that used in 802.11g-b are able to adequately propagate the errors at their input. So, they are ineffective in the proposed physical layer security scheme. In the next section, a new proposal is presented that, starting from the structures in Fig. \ref{fig:scrambler_b}, modifies them in such a way as to obtain the desired properties.

\section{Proposed scrambler}
\label{sec:three}
\begin{figure}[tb]
\begin{centering}
\includegraphics[width=70mm,keepaspectratio]{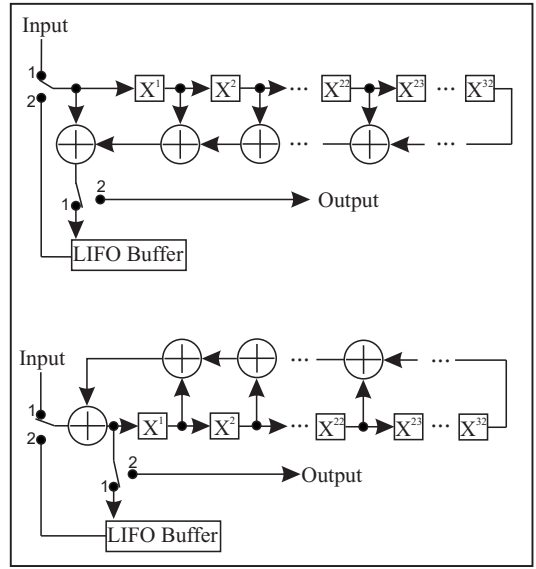}
\caption{Proposed scrambler (top) and descrambler (bottom) circuits.
\label{fig:scrambler32}}
\par\end{centering}
\end{figure}

Let us remind that the goal of the descrambler, in the proposed physical layer security scheme, is to propagate the residual errors, in such a way that even with a single error at its input, about half of the bits at the output are in error.
This result cannot be achieved starting from the IEEE 802.11g-a scrambler/descrambler, while the circuits used in the IEEE 802.11g-b standard can reach this target, through a suitable modification.
The proposed solution is shown in Fig. \ref{fig:scrambler32}.
The main changes, discussed next, are the following:
\begin{itemize}
\item scrambler and descrambler are interchanged;
\item scrambling and descrambling are realized in two steps;
\item the 7-cell LFSR is replaced by one with 32 cells.
\end{itemize}

\begin{figure}[tb]
\begin{centering}
\includegraphics[width=70mm,keepaspectratio]{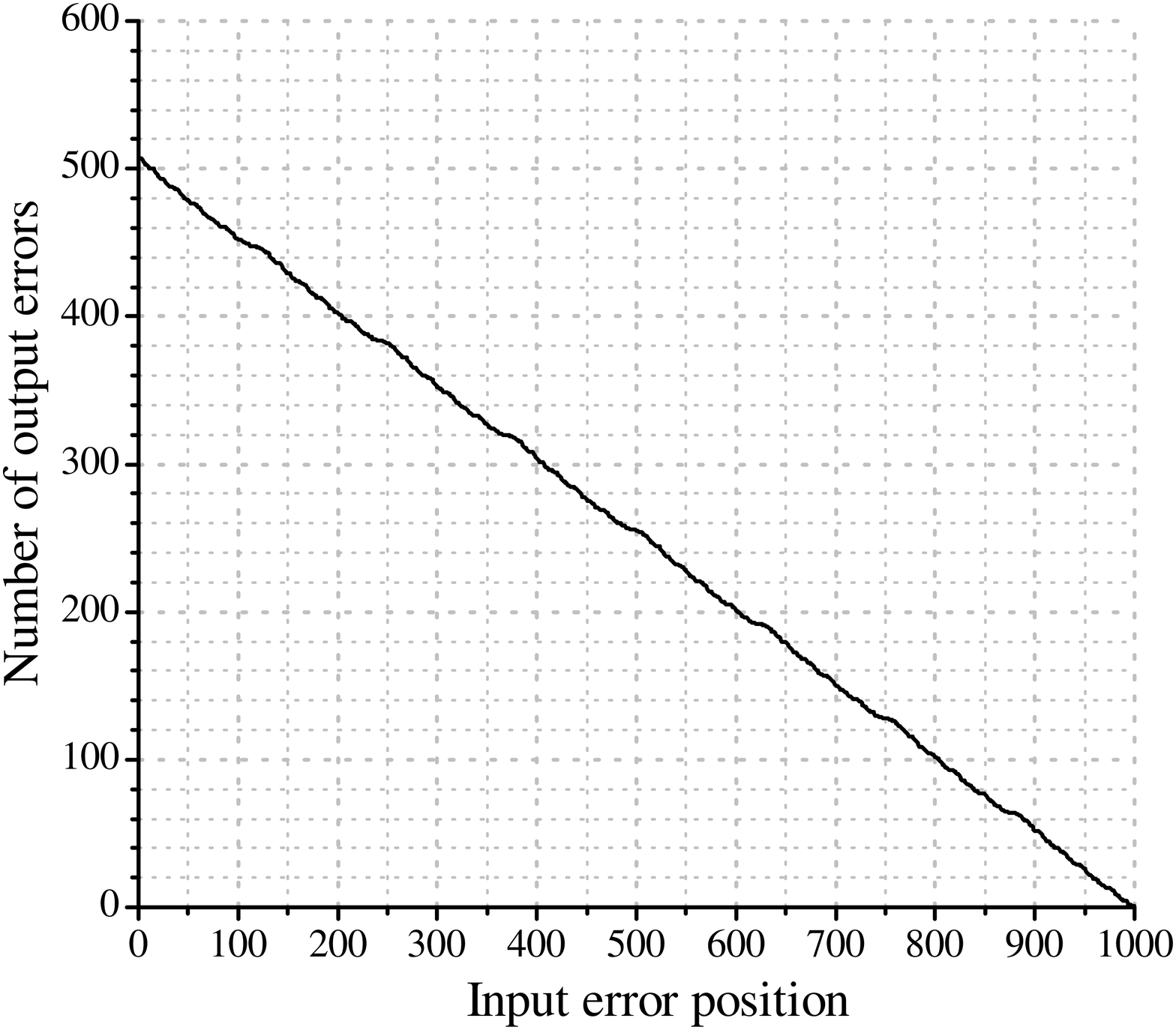}
\caption{Number of output errors versus the position of a single input error, for the IEEE 802.11g-b scrambler with a 1000-bit input  sequence (standard 7-cell LFSR). \label{fig:contrario}}
\par\end{centering}
\end{figure}

{\it Scrambler/descrambler interchange:} The main reason why the 802.11g-b descrambler has no ability to propagate the errors is that it does not include any feedback link able to bring an error back into the LFSR after crossing the seven cells it consists of. On the contrary, as evident in Fig. \ref{fig:scrambler_b}, such property is present in the scrambler circuit. So, the error propagation ability can be improved by interchanging the role of the two circuits.
It is possible to verify that this does not alter the behavior from the synchronization standpoint, hence the standard requirement is not affected by this change.
However, since the standard circuit works on its input in a sequential way, it is able to propagate the errors only on the bits which are input after the first erred one.
This is shown in Fig. \ref{fig:contrario}, where a 1000-bit sequence has been given as input to the 802.11g-b scrambler, with a single erred bit in a variable position.
When the input error is in the initial positions, the scrambler has a good error propagation characteristic; this performance, however, disappears when the error occurs later.

\begin{figure}[tb]
\begin{centering}
\includegraphics[width=70mm,keepaspectratio]{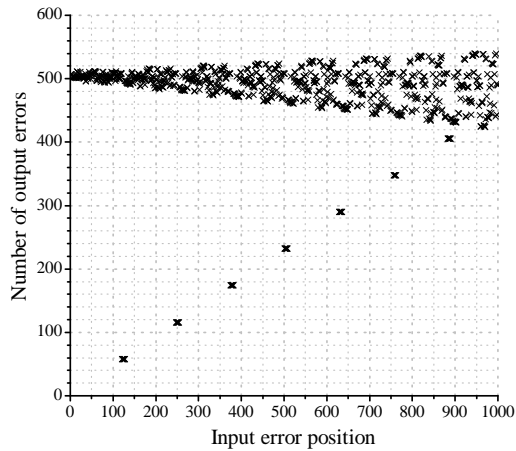}
\caption{Number of output errors versus the position of a single input error, for a two-steps descrambler with 7-cell LFSR and 1000-bit input sequence. \label{fig:contrario_doppio_7}}
\par\end{centering}
\end{figure}

{\it Two-steps scrambling/descrambling:} A simple way to overcome the above limit consists in performing the scrambling and descrambling operations in two steps, first entering the scrambler (descrambler) according to the input order and then in the reverse order. Looking at Fig. \ref{fig:scrambler32}, the switch stays in position 1 for the time necessary to enter the input sequence. The output of the LFSR is stored in the last-in first-out (LIFO) buffer. Then the switch is moved to position 2, and the contents of the buffer are input again to the LFSR.
This way, an error occupying a late position in the first step translates into an early position in the second step, and vice versa.
If we use the standard 7-cell LFSR in the circuit of Fig. \ref{fig:scrambler32},
with a 1000-bit input sequence, we obtain a number of output errors as shown in Fig. \ref{fig:contrario_doppio_7}: in spite of a small dispersion, dependent on the input error position, the target of having an average number of errors at the output close to half the input size is reached with good approximation. This result is further improved next.

\begin{figure}[tb]
\begin{centering}
\includegraphics[width=70mm,keepaspectratio]{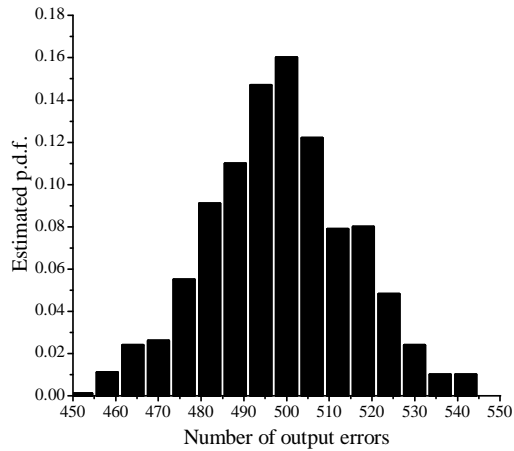}
\caption{Number of output errors distribution for the new scrambler.
\label{fig:new_scrambler_result}}
\par\end{centering}
\end{figure}

{\it 32-cell LFSR:} Fig. \ref{fig:contrario_doppio_7} shows that there are some ``unlucky'' positions where a single error at the descrambler input produces a number of errors at the output significantly smaller than half the input size. These positions are multiple of $127$, i.e., the standard LFSR period, and are due to the device structure.
The way to eliminate such singular points consists in exploiting longer LFSRs.
Taking into account the need to transmit long sequences, which we will discuss afterward, we propose to adopt a 32-cell LFSR, with $G(X) = X^{32} + X^{22} + X^{2} + X + 1$.
The simulated distribution of the output errors for the circuits shown in Fig. \ref{fig:scrambler32} is reported in Fig. \ref{fig:new_scrambler_result}: the probability density function (p.d.f.) is centered at $500$, as expected, and exhibits a limited dispersion around the mean. We have verified that such behavior is independent of the input error position and it is maintained also in the case of multiple errors at the descrambler input.

\section{Probability of correct reception}
\label{sec:four}
Once having demonstrated that the new descrambler is able to properly propagate the errors, we need to verify that the error rate due to the channel is sufficiently high to ensure that the probability of information interception by Eve is very low in practical operation conditions. For such purpose, the traffic through the channel must be characterized. Experimental and simulated results can be used for this purpose.

First of all, by capturing and analyzing the traffic of a real WLAN through common tools like Kismet \cite{b0017} and Wireshark \cite{b0018}, we have evaluated the burstiness level of the packet errors, through the approach presented in \cite{b0002}.
It exploits a metric, called $\beta$ and defined as $\beta = \left[ KW(I) - KW(E) \right]/KW(I)$, where $KW$ is the Kantorovich-Wasserstein distance, that is used to measure how close a conditional packet delivery function (CPDF) is to that of the ideal bursty link. $E$ is the CPDF of the empirical link, while $I$ is the CPDF of an independent link with the same packet reception ratio. A high $\beta$ means the link is very bursty, while a $\beta$ close to zero means the link is independent. Further details can be found in \cite{b0002}.
It is known that many but not all link layers observe burstiness. In our experiments, performed in a common indoor environment, with no retransmissions allowed, we have found $\beta \approx 0.16$, which permits us to consider the errors almost uncorrelated.

\begin{figure}[tb]
\begin{centering}
\includegraphics[width=70mm,keepaspectratio]{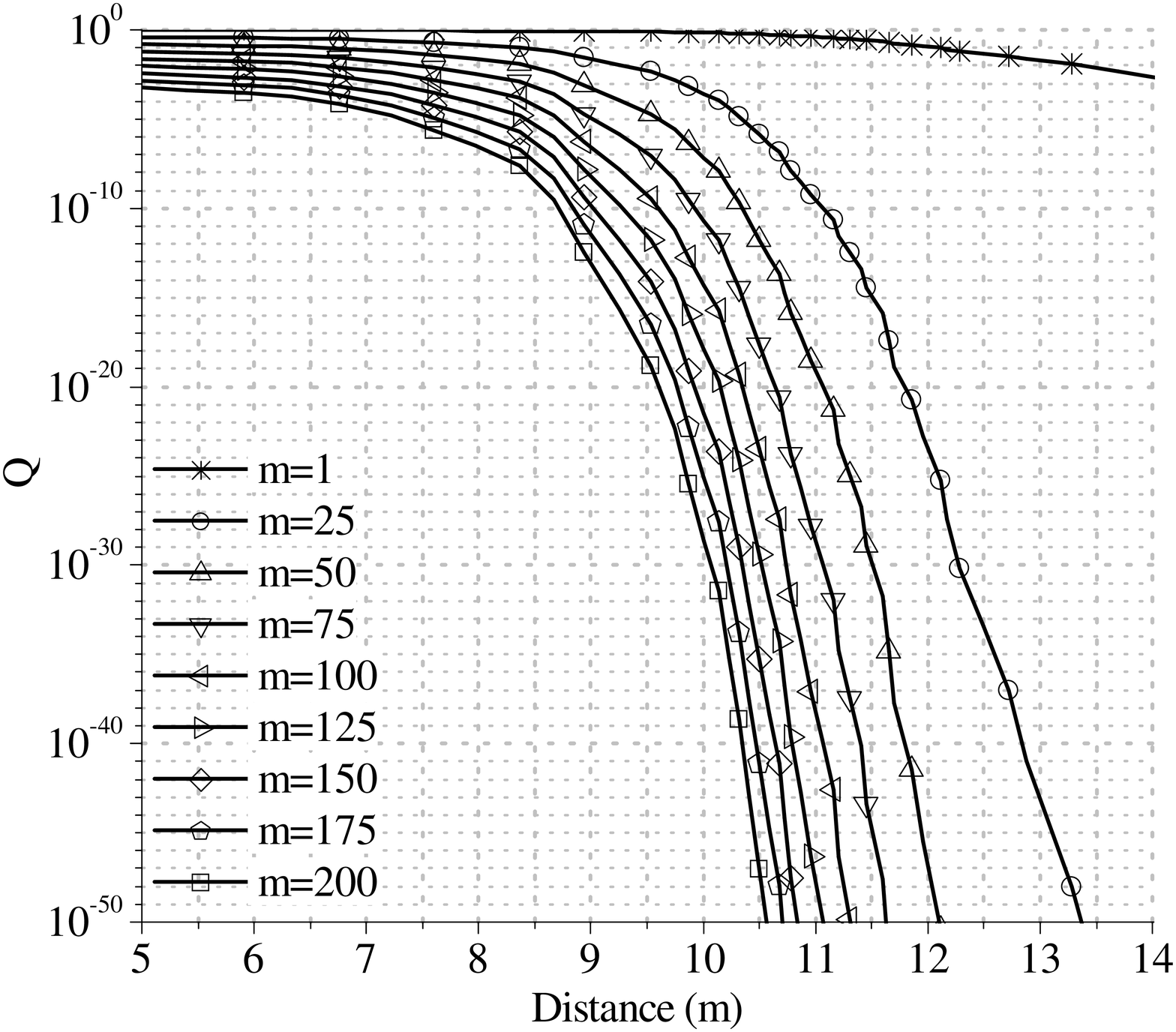}
\caption{Probability of correct reception, as a function of the distance, in a $54$ Mbps WLAN with 200 data bytes packets and $S_T = 6$ dBm.
\label{fig:conc_pl}}
\par\end{centering}
\end{figure}

Denoting by $P$ the channel packet error rate, the probability that a block of $m$ packets is received without errors, under the hypothesis of an independent channel, is
\begin{equation}
Q = (1 - P)^m. 
\label{eq:Q}
\end{equation}
The value of $Q$ also depends on the transmitted power $S_T$. Based on the path loss model presented in \cite[Eq. (2)]{Seidel1992} and the values of $P$ reported in \cite{b0007} [where packets with $L = 200$ bytes of application data and $S_T = 12$ dBm are considered] we have used \eqref{eq:Q} to obtain a set of curves of $Q$ as a function of the distance from the AP. These curves are shown in Fig. \ref{fig:conc_pl}, for the case of a $54$ Mbps data rate, $S_T = 6$ dBm and several values of $m$. Similar behaviors can be obtained for the other data rates of the standard. These curves can be used for either Bob or Eve: for the former, the distance should be sufficiently small to have a large $Q$ (higher than a prefixed threshold $Q_B$); for the latter, the distance should be sufficiently large to have a small $Q$ (lower than a given threshold $Q_E$). For the considered example, in particular, we see that by assuming $m = 125$, the unauthorized user has $Q < Q_E = 10^{-30}$ for a distance of $10.5$ m or more. A relevant point, emerging from the figure, is that the slope of the curves can be made very steep for values of $m$ sufficiently large; this means that the transition from the secure region to the zone where Eve can stand without endangering security can be made very sharp, thus reaching the target of the proposed physical layer system.
We remark that, for this purpose, the $m$ packets must be scrambled consecutively, without resetting the scrambler (and descrambler) at the beginning of each of them.

An apparent drawback of the proposed system is the time required for key acquisition, i.e., the secure WLAN bootstrap time ($T_B$). This can be expressed as
\begin{equation}
T_B = (mL/R_b + T_r)/Q,
\label{eq:TB}
\end{equation}
where $R_b$ is the bit rate and $T_r$ the repetition period for the key transmission. Because of the high value of $R_b$ (e.g., $R_b = 54$ Mbps in Fig. \ref{fig:conc_pl}) the first contribution is always negligible and $T_B \approx T_r/Q$. Thus, by assuming, as an example, $Q_B = 10^{-1}$ and $T_r = 1$ s, $T_B$ is in the order of $10$ s. This value is not negligible; however, it should be noted that the network bootstrap must be done only once for each session, so that, in relative terms, the bootstrap time is quite acceptable.

\section{Conclusion}
\label{sec:five}
We have proposed a physical layer security technique to implement the automatic bootstrap and renewing of the network key in IEEE 802.11g WLANs.
The proposed solution exploits a suitable scrambler/descrambler apparatus, which is able to approach the error propagation performance of a perfect scrambler, while preserving the flexibility and simplicity of the original IEEE 802.11 standard circuits.
So, it opens the possibility to implement, at a very limited cost, effective mechanisms of physical layer security in worldwide used wireless networks. 



\end{document}